\documentclass[
english,aps,pre
]{revtex4}
\usepackage{babel}
\usepackage{graphics}

\makeatletter

\providecommand{\LyX}{L\kern-.1667em\lower.25em\hbox{Y}\kern-.125emX\@}
\let\SF@@footnote\footnote
\def\footnote{\ifx\protect\@typeset@protect
    \expandafter\SF@@footnote
  \else
    \expandafter\SF@gobble@opt
  \fi
}
\expandafter\def\csname SF@gobble@opt \endcsname{\@ifnextchar[
  \SF@gobble@twobracket
  \@gobble
}
\edef\SF@gobble@opt{\noexpand\protect
  \expandafter\noexpand\csname SF@gobble@opt \endcsname}
\def\SF@gobble@twobracket[#1]#2{}

\usepackage{graphicx}
\usepackage{bm}

\makeatother
\begin{document}

\title{Thermodynamics of the self-gravitating ring model}
\author{Takayuki Tatekawa$^{1,2}$\thanks{E-mail: tatekawa@gravity.phys.waseda.ac.jp},
  Freddy Bouchet$^{1}$\thanks{E-mail: Freddy.Bouchet@ens-lyon.fr}, Thierry
  Dauxois$^{1}$\thanks{E-mail: Thierry.Dauxois@ens-lyon.fr}, Stefano
  Ruffo$^{1,3}$\thanks{E-mail: ruffo@avanzi.de.unifi.it} }
\affiliation{1. Laboratoire de Physique, UMR-CNRS 5672, ENS Lyon,
46
  All\'{e}e d'Italie, 69364 Lyon c\'{e}dex 07, France\\
  2. Department of Physics, Waseda University, 3-4-1 Okubo,
Shinjuku, Tokyo, 169-8555, Japan\\
 3. Dipartimento di Energetica, ``S. Stecco'' and CSDC, Universit{\`a} di
  Firenze, and INFN, via S. Marta, 3, 50139 Firenze, Italy }
\date{\today}

\begin{abstract}
We present the phase diagram, in both the microcanonical and the canonical
ensemble, of the Self-Gravitating-Ring (SGR) model, which describes the
motion of equal point masses constrained on a ring and subject to 
3D gravitational attraction. If the interaction is regularized at
short distances by the introduction of a softening parameter, a 
global entropy maximum always exists, and thermodynamics
is well defined in the mean-field limit. However, {\it ensembles are 
not equivalent} and a phase of {\it negative specific heat} in the 
microcanonical ensemble appears in a wide intermediate energy 
region, if the softening parameter is small enough. The phase 
transition changes from second to first order at a {\it tricritical} 
point, whose location is not the same in the two ensembles.
All these features make of the SGR model the best prototype 
of a self-gravitating system in one dimension. In order to obtain
the stable stationary mass distribution, we apply a new iterative method,
inspired by a previous one used in 2D turbulence, which ensures entropy 
increase and, hence, convergence towards an equilibrium state. 
\end{abstract}


\maketitle

\section{Introduction}\label{sec:intro}

There are many objects in our universe whose behavior can be understood
considering only the gravitational interaction. Examples are globular clusters,
galaxies, clusters of galaxies, molecular clouds~\cite{BT}.
Different theoretical approaches have been proposed to explain
the peculiar  statistical properties of self-gravitating systems.
The main difficulty is that these systems cannot approach statistical
equilibrium because of the short-distance divergence of the potential
and of the evaporation  at the boundaries. Even if one puts the system in a box
with adiabatic walls, thus eliminating evaporation, still gravity causes
the well-known phenomenon of
{\em gravothermal catastrophe}~\cite{Antonov, Lynden-bell, Padmanabhan89}.
The introduction of a small-scale softening of the interaction potential
avoids such a catastrophe, so that self-gravitating systems
can approach the final (thermal) equilibrium state.
However, such a state may have a {\em negative specific heat}.
Moreover, a
first-order phase transition from the high energy gas phase to
the low energy clustered phase appears~\cite{Lynden-bell}.

Direct studies of the full three-dimensional $N$-body gravitational dynamics
are particularly heavy~\cite{Heggiehut} and even special purpose machines
have been built to this aim~\cite{grape5taka}. Therefore, lower dimensional models
have been introduced to describe gravitational systems with additional symmetries.
For instance, the gravitational sheet model describing the motion of infinite planar mass distributions
perpendicularly to their surface has been considered~\cite{Feix}.
Although this model shows interesting behaviors~\cite{Tsuchiya94,Koyama01},
the specific heat is always positive and no phase transition is present.

Recently, another one-dimensional model has been introduced~\cite{SGR-model}
where particle motion is confined on a ring, but the interaction is the true Newtonian
3D one. At short distances, the potential is regularized,
so that the particles do not interact. This model has been called the
Self-Gravitating Ring model (SGR) and will be  the subject of the study discussed
in this paper. It has been shown in numerical simulations~\cite{SGR-model},
that this model maintains the peculiar features of the 3D Newtonian potential,
showing a negative specific heat phase and a phase transition if the softening parameter is small enough.
Moreover, for large softening, this model reduces to the Hamiltonian Mean-Field model (HMF)~\cite{HMF-model},
which has been recently extensively studied as a prototype system with long-range interactions.
This latter model, however, although it displays a second order phase transition, 
does not have a negative specific heat phase at equilibrium.

In this paper, we derive the equilibrium thermodynamics of
the SGR model both in the canonical and in the microcanonical ensemble.
For all non vanishing softening parameter values,
this model has a thermal equilibrium state. If the softening parameter is
small enough, the model shows {\em ensemble inequivalence}~\cite{Dauxois02LNP,Classification}
with a negative specific heat phase in the microcanonical ensemble and a first order
phase transition.
Therefore, the SGR model displays several features of the true 3D
Newtonian interaction, and can serve as a better prototype of self-gravitating systems
in one dimension than all previously introduced models.

The paper is organized as follows.
In Sec.~\ref{sec:SGR-model}, we briefly introduce the SGR model
and we discuss the essential features of previous numerical simulations~\cite{SGR-model}.
In Sec.~\ref{sec:Stable-dist}, we show the general scheme for deriving
all stationary density distributions which maximize Boltzmann-Gibbs
entropy at fixed total energy and mass.
Section~\ref{sec:Iterative} presents a new iterative method
which ensures entropy increase and leads in a unique way towards the stable equilibrium single particle
distribution function. The method is inspired by a similar one
used to compute entropy maxima in 2D turbulence~\cite{Turkington}.
In Sec.~\ref{implementaionalgorithm}, we describe in full detail how to implement
the iterative algorithm in a numerical scheme.
In Sec.~\ref{sec:NumericalSol}, we calculate the thermodynamic
quantities of the SGR model using the iterative method. We also show
 how, reducing the softening parameter, one enters into a region of ensemble inequivalence,
where a tricritical point exists which is not the same in the two ensembles~\cite{Barre_Mukamel_Ruffo}.
Finally, in Sec.~\ref{sec:Evolution}, we discuss the dynamical evolution
of the SGR model, emphasizing the properties of relaxation to equilibrium.

\section{Self-gravitating ring model}\label{sec:SGR-model}

\begin{figure}
 \resizebox{60mm}{!}{
 \includegraphics{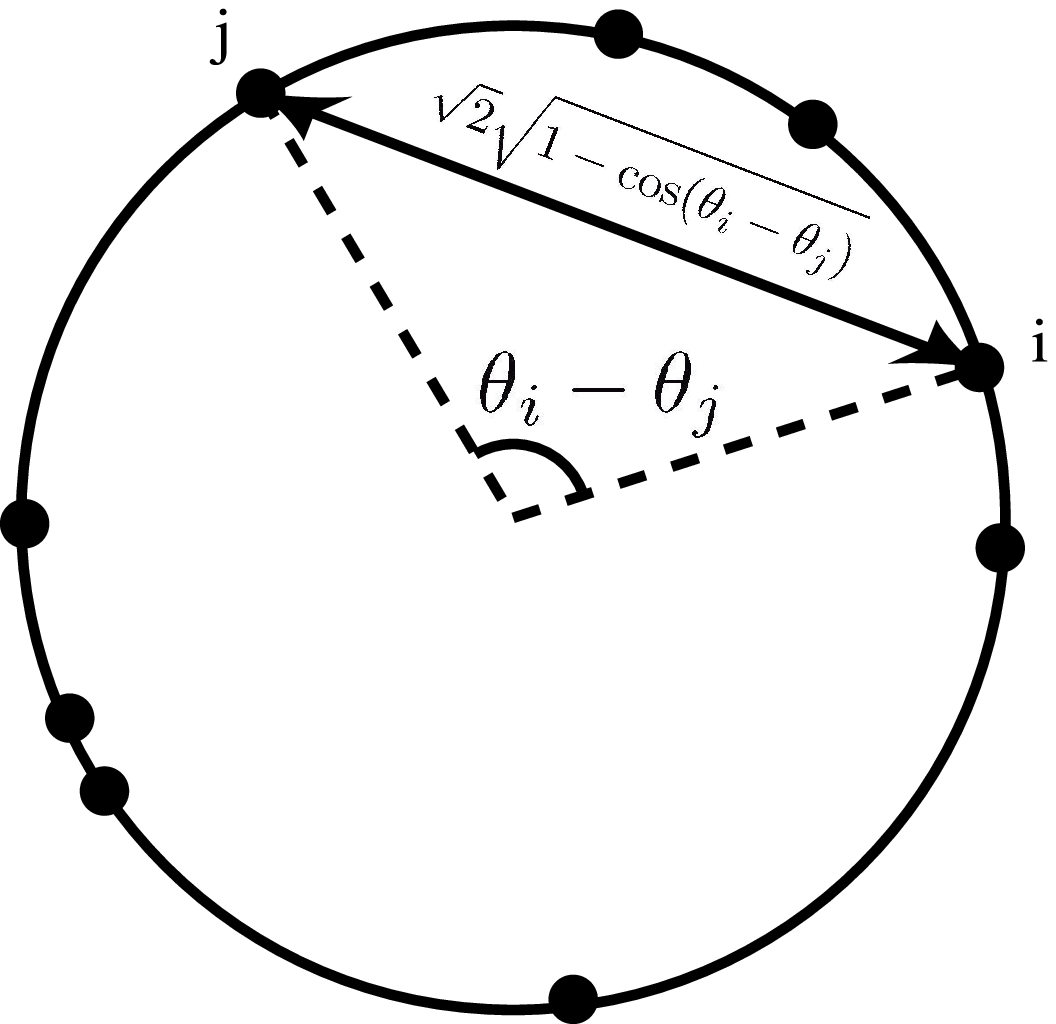}
 }
 \caption{
   Self-Gravitating Ring model with a fixed unitary radius.
   Particles are constrained to move on a ring and therefore their
   location is specified by the angles measured with respect to a
   fixed direction. Each pair of particles at $\theta_i$ and $\theta_j$
   interacts through the inverse-square three-dimensional
   gravitational force.  The distance is measured by the chord, as
   shown  in the figure.}
\label{fig:SGR-figure}
\end{figure}

In this section, we briefly present the Self-Gravitating Ring (SGR)
model~\cite{SGR-model}. In this model, particle motion
is constrained on a ring  and particles interact via
a true 3D Newtonian potential (Fig.~\ref{fig:SGR-figure}).

The Hamiltonian of the SGR model is
\begin{eqnarray} \label{eqn:SGR-H}
H &=& \frac{1}{2} \sum_{i=1}^N p_i^2 + \frac{1}{2N} \sum_{i,j}
 V_{\varepsilon} (\theta_i - \theta_j) \,, \\
V_{\varepsilon} (\theta_i - \theta_j)& = & -\frac{1}{\sqrt{2}}
 \frac{1}{\sqrt{1-\cos (\theta_i - \theta_j) + \varepsilon}} \,,
\end{eqnarray}
where $\varepsilon$ is the softening parameter, which is introduced,
as usual, in order to avoid the divergence of the potential at short distances.

\begin{figure}
\resizebox{80mm}{!}{ \includegraphics{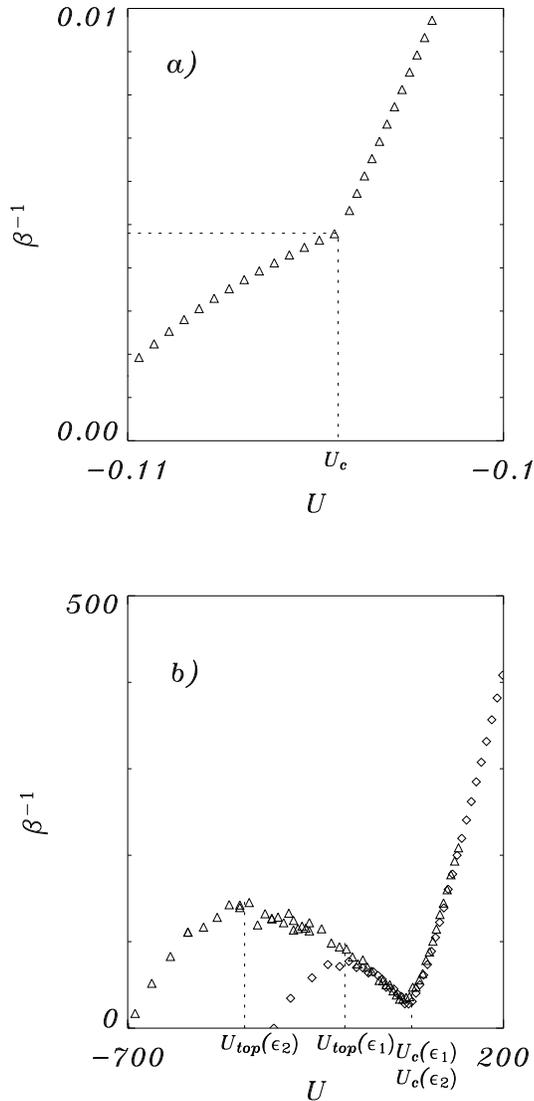} }
 \caption{Caloric curves of the Self Gravitating Ring (SGR) model obtained from numerical simulations
of Hamiltonian~\ref{eqn:SGR-H}.
   Panel (a) refers to the softening parameter value $\varepsilon=10$, for which
   a second order phase transition appears at $U_c$. No backbending of the caloric curve,
indicating a negative specific heat, is present. Simulations were performed for
$N=100$. Panel (b) shows the caloric curves for two different values
   of the softening parameter, $\varepsilon_1=1.0 × 10^{-6}$ and $\varepsilon_2= 2.5 × 10^{-7}$, and $N=100$.  The
   transition is here first order in the microcanonical ensemble (see
   Sec.~\ref{sec:NumericalSol} for a discussion). The two transition energies
   $U_{c} (\varepsilon_1)$ and $U_{c} (\varepsilon_2)$ are pretty close, suggesting a slow
   variation of the critical energy with the softening parameter $\varepsilon$.
   On the contrary, $U_{top} (\varepsilon_1)$ is significantly smaller than $U_{top}
   (\varepsilon_2)$, indicating that this characteristic energy value diminishes
   with $\varepsilon$. A negative specific heat phase appears for $U_{top}< U< U_c$,
and expands as the softening parameter is reduced.}
\label{fig:SGR-caloric}
\end{figure}

Taking the  large $\varepsilon$ limit, the potential becomes
\begin{equation} \label{eqn:large-e-limit}
V_{\varepsilon}= \frac{1}{\sqrt{2 \varepsilon} } \left[\frac{1-\cos \left(\theta_i - \theta_j\right)}{2\varepsilon}-1
 \right]  + O(\varepsilon^{-2}) \,,
\end{equation}
which is the one of  the Hamiltonian Mean-Field (HMF) model~\cite{HMF-model}.
It is well known that the HMF model~\cite{HMF-model} has a second
order phase transition, separating a low energy phase, where the particles
form a single cluster, from a high energy gas phase where kinetic
energy dominates and the particles are homogeneously distributed on
the circle. One usually draws the so-called caloric curve, where
temperature, given by twice the averaged kinetic energy per particle $T\equiv\beta^{-1}={2
  \left<K \right>}/{N}$, is plotted against the total energy per
particle $U\equiv{H}/{N}$. In a situation close to that of the HMF model,
e.g. for $\varepsilon=10$, the caloric curve determined from microcanonical
numerical simulations is reported in Fig.~\ref{fig:SGR-caloric}(a).
In the homogeneous phase $U>U_c(\varepsilon)$, the caloric curve is almost linear, while
in the clustered phase $U<U_c(\varepsilon)$, it is bent downward.  Nonetheless,
temperature always grows with energy and one does not observe any
negative specific heat energy range. However, as it happens for 3D
Newtonian gravity simulations~\cite{Heggiehut}, when one reduces the softening
parameter, a negative specific heat phase develops. For instance, in
Fig.~\ref{fig:SGR-caloric}(b), we show two cases at small
$\varepsilon$ where three phases can be identified~\cite{SGR-model}:
\begin{itemize}
\item a low-energy clustered phase  for $U<U_{top} (\varepsilon)$,
  where $U_{top}$ is defined as the energy at which $\partial T/\partial U=0$.
\item an intermediate-energy phase,
  $U_{top}(\varepsilon)<U<U_c(\varepsilon)$, with negative specific heat.
\item a high-energy gaseous phase for  $U_{c}(\varepsilon) <U$.
\end{itemize}
The clustered phase is created by the presence of softening $\varepsilon$,
without which the particles would fall into the zero distance
singularity.  In the gas phase, the particles are hardly
affected by the potential and behave as almost free particles.  The
intermediate phase is expected to show the characters of
gravity, persisting and even widening in the $\varepsilon\to0$ limit.

In the following, several of these features will be given a theoretical
explanation and we will detail the analysis of the nature
of the phase transition (first or second order) when $\varepsilon$ is varied.

\section{Stationary density distribution}
\label{sec:Stable-dist}

In the mean-field limit ($N\to\infty $ with fixed length~\cite{spohn}), one can
introduce the single particle distribution function
$f(p, \theta)$ such that $f(p, \theta){\rm d}p {\rm d}\theta $ is the fraction of particles
in the domain $[\theta,\theta+{\rm d}\theta]×[p,p+{\rm d}p]$. In terms of $f$, the
 potential energy can be written  as
\begin{eqnarray}
E_P[f] &=& \frac{1}{2} \int {\rm d} \theta ~{\rm d} \phi
  ~{\rm d}p ~{\rm d} p' ~
 f(p, \theta) V_{\varepsilon} (\theta-\phi) f(p', \phi) \\
&=& \frac{1}{2} \int {\rm d} \phi ~{\rm d} \theta
 ~\rho(\theta) \rho(\phi) V_{\varepsilon} (\theta-\phi)\label{poteneergy}
\end{eqnarray}
where
\begin{equation}
\rho(\theta) = \int {\rm d} p ~f(p, \theta)
\end{equation}
is the mass density.  The kinetic energy is
\begin{equation}
E_K[f] = \frac{1}{2} \int {\rm d} \theta  ~{\rm d}p ~ p^2 f(p, \theta) \label{kineneergy}
\end{equation}
and the total energy
\begin{equation}
E[f]= E_K[f]+E_P[f].
\end{equation}

The equilibrium distribution in the microcanonical
ensemble is determined by maximizing
entropy
\begin{equation} \label{eqn:entropy}
S[f] = -\int {\rm d} \theta ~{\rm d} p ~f \log f
\end{equation}
under the constraints of fixed total energy, momentum and mass.
In the following, we fix the total energy
$E [f] = U $, the total mass
\begin{equation}M [f]=\int \rho  ~ {\rm d} \theta = 1\label{eqn:total-mass}\end{equation}
and the total momentum
\begin{equation}
p[f] =\int p f(p, \theta)
{\rm d} \theta ~{\rm d} p = 0.
\end{equation}

A necessary condition to get an entropy maximum is to require that the free energy
\begin{equation} \label{eqn:entropy+L}
F[f] \equiv S[f] - \beta E[f] - \alpha \int f ~{\rm d} p
 ~{\rm d} \theta - \gamma p[f] ,
\end{equation}
where $\alpha$, $\beta$ and $\gamma$ are Lagrange multipliers,
is stationary
\begin{equation} \label{eqn:delta-entropy+L}
\frac{\delta F[f]}{\delta f}=
-\log f -1 - \beta \left(\frac{p^2}{2} + W(\theta) \right ) - \alpha
 - \gamma p =0 ,
\end{equation}
where $W(\theta)$ is defined as
\begin{equation} \label{eqn:definition-W}
W(\theta) \equiv \int_{-\pi}^{+\pi} \rho(\phi) V_{\varepsilon} (\theta-\phi)
{\rm d} \phi .
\end{equation}

Since $p[f]=0$, the Lagrange multiplier $\gamma$ vanishes.  From
Eq.~(\ref{eqn:delta-entropy+L}), the normalized stationary distribution function can be
written as
\begin{equation} \label{eqn:dist-fn}
f(p, \theta) = A\exp \left [ -\beta \left (\frac{p^2}{2} + W(\theta)
\right ) \right ] ,
\end{equation}
where $A=\exp(-1-\alpha)$ is the normalization constant
and the mass density is given by
\begin{equation} \label{eqn:stable-rho}
\rho(\theta)=\widetilde{A}\ e^{-\beta W(\theta)},
\end{equation}
where $\widetilde{A}=A \sqrt{{2\pi}/{\beta}}$.
When Eq.~(\ref{eqn:definition-W}) and~(\ref{eqn:stable-rho}) are
combined, we obtain the consistency  equations
\begin{eqnarray} \label{eqn:LE-1}
W(\theta) &=& {\widetilde{A}}{}\ \int_{-\pi}^{+\pi} e^{-\beta W(\phi)}
 V_{\varepsilon} (\theta-\phi) {\rm d} \phi ,
\end{eqnarray}
and the equilibrium density equation
\begin{eqnarray}
\rho(\theta) &= & \widetilde{A}\exp \left[ -{\beta\widetilde{A}} \int_{-\pi}^{+\pi}
 \rho(\phi) V_{\varepsilon} (\theta-\phi) {\rm d} \phi \right ] , \label{eqn:LE-2}
\end{eqnarray}
which are solved numerically in the following.  Once the stationary
mass distributions $\rho$ and the function $W$ are obtained for each value of
$\varepsilon$, the full single particle distribution function $f(\theta,p)$ is
derived from Eq.~(\ref{eqn:dist-fn}). The potential energy and the
kinetic energy are determined by Eq.~(\ref{poteneergy}) and
Eq.~(\ref{kineneergy}) respectively, allowing to draw the caloric
curve by plotting $T\equiv\beta^{-1}=2E_K$ against the total energy $U=E_K+E_P$.

\section{An iterative method to solve the   equilibrium density  equation}
\label{sec:Iterative}

The inverse temperature $\beta$ can be expressed in terms of the energy
\begin{equation} \label{eqn:beta-E-V}
\beta= \left \{ 2U -  \int_{-\pi}^{+\pi} \int_{-\pi}^{+\pi}
 \rho(\theta) \rho(\phi) V_{\varepsilon} (\theta-\phi)
 {\rm d} \theta {\rm d} \phi \right \}^{-1} .
\end{equation}
Once an initial density distribution $\rho_0(\theta)$ is chosen,
one can determine an initial inverse temperature $\beta_0$ using
Eq.~(\ref{eqn:beta-E-V}), and then solve the consistency
equation~(\ref{eqn:LE-2}) iteratively (as done for instance in
Ref.~\cite{ispolatovcohen}).  However, we will follow here a different
iterative method, which ensures entropy increase and, hence,
convergence of the algorithm.  The method is inspired by a similar one
used by Turkington and Whittaker~\cite{Turkington} to compute entropy
maxima for two dimensional turbulence.

The functional to maximize $S[f]$ is strictly concave
and we must fix both a linear constraint $M[f]=1$
and a nonlinear one $ E[f] =U$. It is this latter nonlinear constraint
which makes the variational problem more difficult than usual.
The trick to solve this nonlinear problem consists in considering a
linearization of the energy constraint around the distribution
function resulting from the previous step in the iterative process.

One begins with the normalized distribution $f_k$ obtained at the
$k$th step of the algorithm. From that, one can compute the mass
density~$\rho _{k}$ and the average potential~$W_{k}$.
\begin{eqnarray}
\label{eqn:ro et W}
\rho _{k}(\theta )&=&\int {\textrm{d}}\,p\,f_k(p,\theta )
\\
W_{k}(\theta )&=&\int _{-\pi }^{+\pi }{\textrm{d}}\phi\, \rho_{k}(\phi )
\,V_{\varepsilon }(\theta -\phi ).\label{eqn:ro et Wbis}
\end{eqnarray}
The distribution at the next step $f_{k+1}$ will be then determined by
solving the following variational problem
\begin{eqnarray}
\label{eqn:variational_iteration}
\mbox{max} \biggl\{ S[f]\ \left|\right. M[f] =1,
   E [f_{k}]+\int \left.\frac{\delta   E}{\delta   f}\right|_{f_{k}}
\left(f-f_{k}\right) {\textrm{d}}p{\textrm{d}}\theta \leq U\biggr\} ,
\end{eqnarray}
where the functional derivative of the energy is
\begin{equation}
 \left.\frac{\delta   E}{\delta   f}\right|_{f_{k}} =\frac{p^{2}}{2}+W_{k}\left( \theta \right).
\end{equation}
This variational problem has a unique solution $f_{k+1}$, since
it corresponds to the maximization of a strictly concave functional
with linear constraints.

This iterative process ensures convergence of the
entropy. Let us prove it. By using a generalization of
the Lagrange multiplier rule for our inequality constrained
variational problem~\cite{IOffe,Rockafellar}
\begin{equation}
\label{eqn:first_variations_iteration}
\left.\frac{\delta S}{\delta   f}\right|_{f_{k+1}} =\alpha _{k+1}
+\beta _{k+1} \left.\frac{\delta   E}{\delta   f}\right|_{f_{k}}
\end{equation}
with the additional requirement
\begin{equation}
\label{eqn:first_variations_iteration_2}
 \beta _{k+1}\left[ E\left[ f^{k}\right] +\int \left.\frac{\delta   E}{\delta
       f}\right|_{f_{k}}  \left( f_{k+1}-f_{k}\right) {\textrm{d}}p{\textrm{d}}
\theta -U\right] =0 ,
\end{equation}
where $ \beta _{k+1}\geq0$  is the multiplier associated with the energy
constraint and $\alpha_{k+1}$, the one associated with mass conservation.
When solving Eq.~(\ref{eqn:first_variations_iteration_2}), we have
either $\beta _{k+1}=0$, which removes the energy constraint, or $\beta
_{k+1}>0$, and an equality for the linearized energy constraint.

In order to prove convergence of the entropy, let us first prove
that the energy functional $E\left[ f\right]$ is concave.
Since the kinetic part is linear in $f$, the second variation of
$E[f]$ is
\begin{eqnarray}
\label{eqn:second_variation_energy}
\delta^{2}E&=&\int {\textrm{d}}\phi {\textrm{d}}\theta\,
\delta \rho (\theta )\,\delta \rho (\phi )\,V_{\varepsilon }(\theta -\phi )\\
&=&\displaystyle
\sum_{k}V_{\varepsilon ,k}\left| \delta \rho _{k}\right| ^{2} ,\label{eqn:second_variation_energy2}
\end{eqnarray}
where the second equality is obtained using the Fourier series
expansion for both the mass density variation
$\delta \rho$ and the potential $V_{\varepsilon }$
\begin{eqnarray}  \delta \rho _{k}&=&\frac{1}{2\pi}
\int ^{+\pi }_{-\pi }{\textrm{d}}\varphi \, \exp \left( ik\varphi \right) \delta \rho
\left( \varphi \right)  \\
V_{\varepsilon ,k}&=&\frac{1}{2\pi}
\int ^{+\pi }_{-\pi }{\textrm{d}}\varphi  \, \exp \left( ik\varphi \right) V_{\varepsilon }
\left( \varphi \right)  .
\end{eqnarray}

Since $V_{\varepsilon }$ is even in the argument~$\phi$, $V_{\varepsilon ,k}$ is a real number.
Moreover, since $V_{\varepsilon }\leq 0$ and $ V_{\varepsilon }\left(\varphi\right)$ is strictly
increasing for $ 0\leq \varphi \leq \pi $, it is easy to prove that for any $k$,
$V_{\varepsilon ,k}$ is strictly negative.  Hence, from
formula~(\ref{eqn:second_variation_energy2}) the second variation of
the energy functional is negative and this functional is strictly
concave.

On the other hand, the entropy is strictly concave. We have
\begin{equation} S[f+\delta f]\leq S[f]+\int{\textrm{d}}\theta {\textrm{d}}p\,
\left.\frac{\delta S}{\delta   f}\right|_{f}
\delta f-\frac{1}{2}\int {\textrm{d}}\theta {\textrm{d}}p\,\frac{\left( \delta f\right)
  ^{2}}{f} ,\end{equation}
where in the derivation we have used $\ln(1+x) \geq x-x^2/2$ for $x>-1$.
Applying this latter inequality with $ f=f_{k+1} $ and $\delta f=f_{k}-f_{k+1}$,
and using both condition~(\ref{eqn:first_variations_iteration})
and~(\ref{eqn:first_variations_iteration_2}), we obtain \begin{equation}
\label{inequality_1}
S\left[ f_{k+1}\right] -S\left[ f_{k}\right] \geq \beta _{k+1}
\left( U-E\left[ f_{k} \right]\right) +\frac{1}{2}\int{\textrm{d}}\theta {\textrm{d}}p\,
\frac{\left( f_{k+1}-f_{k}\right) ^{2}}{f_k},
\end{equation}
where the term proportional to $\alpha_{k+1}$ vanishes because of mass
conservation.

We will now use the concavity of the energy functional
$E\left[ f\right]$. For $ k>1 $,
\begin{equation} \mathrm{E}\left[ f_{k}\right] \leq
E\left[ f_{k-1}\right] +\int  \left.\frac{\delta E}{\delta   f}\right|_{f_{k-1}}
\left( f_{k}-f_{k-1}\right) {\textrm{d}}p{\textrm{d}}\theta.  \end{equation}

As $\beta_{k+1}\geq 0$ and $\mathrm{E}\left[ f_{k}\right] \leq U$, directly
from the variational problem~(\ref{eqn:variational_iteration}),
Eq.~(\ref{inequality_1}) implies that
\begin{equation}
\label{inequality_2b}
 S\left[ f_{k+1}\right] -S\left[ f_{k}\right]
\geq   \frac{1}{2}\int \frac{\left(
    f_{k+1}-f_{k}\right)^{2}}{f_k}{\textrm{d}}\theta {\textrm{d}}p\geq 0.
\end{equation}
Hence, the entropy has to increase for all iterates after the second.
Since the entropy is bounded from above, it has to converge.
Using Eqs.~(\ref{inequality_1}) and~(\ref{inequality_2b}), one derives
that  the
energy $ \mathrm{E} \left[ f_{k}\right]$ converges to $U$ from below.
Moreover, assuming that $f_{k}$ converges toward $f$, one can prove
the convergence of the multipliers to limit values $\alpha$ and $\beta\geq0$,
which implies that $f$ verifies
Eq.~(\ref{eqn:first_variations_iteration}) for equilibrium states.
Although mathematically one cannot prove the convergence of $f$,
in all practical cases we will analyze, it appears to be verified.
For a more thorough discussion of the convergence in the similar case 
of the Euler equation, see Sec. IV in
Ref.~\cite{Turkington}.


\section{Implementation of the algorithm}
\label{implementaionalgorithm}

We describe in this section the practical implementation of
an algorithm which allows the calculation of the stable distribution,
using the method described in the previous section.

From~(\ref{eqn:first_variations_iteration}), we obtain
\begin{equation}
\label{eqn:fkplusun}
f_{k+1}=A_{k+1}\exp \left( -\beta _{k+1}\left( \frac{p^{2}}{2}+W_{k}
\left( \theta \right) \right) \right) ,\end{equation}
where $A_{k+1}=\exp(-\alpha _{k+1}-1)$ and $\beta_{k+1}$
are unknown at this stage.
Using~(\ref{eqn:ro et W}), we get
\begin{equation}\label{equationpourrhok}
\rho _{k+1}(\theta )={\widetilde{A}}_{k+1}\,e^{-\beta _{k+1}W_{k}(\theta )}\, .
\end{equation}
where ${\widetilde{A}}_{k+1}=A_{k+1}\sqrt{2\pi/\beta_{k+1}}$.  This equation
allows to compute $W_{k+1}(\theta)$ from Eq.~(\ref{eqn:ro et Wbis}) and
\begin{eqnarray} \label{eqn:Func-Wk}
E_{k+1}\equiv E\left[ f_{k+1}\right] & = & \frac{1}{2\beta_{k+1}}+\frac{1}{2}\int _{-\pi }^{+\pi }
 \rho_{k+1}(\theta )W_{k+1}(\theta ){\textrm{d}}\theta  .
\end{eqnarray}
Then the multipliers $\alpha_{k+1}$ and $\beta _{k+1}$ must be computed from
Eqs~(\ref{eqn:total-mass})
and~(\ref{eqn:first_variations_iteration_2}) and, from these, one
gets ${\widetilde{A}}_{k+1}$.  In order to  compute numerically these Lagrange
parameters, let us define the Lagrangian~\cite{Rockafellar}
\begin{eqnarray}
L_k[f](\beta,\alpha) = - S[f] + \beta \left[ E_k + \int \left.\frac{\delta   E}{\delta
       f}\right|_{f_{k}}  \left( f - f_{k}\right) {\textrm{d}}p{\textrm{d}}
\theta -U \right] + \alpha (M[f]-1).
\end{eqnarray}
From this, one further defines
\begin{equation}
L_k^\star(\beta,\alpha) = \inf_f \left\{ L_k[f](\beta,\alpha) \right\}.
\end{equation}
One can prove on a
general ground~\cite{Rockafellar} that $L_k^\star$ is concave and that ${\alpha}_{k+1}$ and $\beta _{k+1}$ are the unique
maxima of $L_k^\star$. Using condition~(\ref{eqn:fkplusun}) for the extrema of $L_k[f](\beta,\alpha)$, we can
compute $L_k^\star$. We obtain, using for practical reasons the variable $\widetilde{A}$ instead of $\alpha$,
\begin{eqnarray} L_{k}^\star(\beta,\widetilde{A}) & = & \log
  \widetilde{A} + \frac{1}{2}\log \beta - \beta \left( U+E_{k}-\frac{1}{2\beta_k} \right) - \widetilde{A}\int  _{-\pi }^{+\pi
    }{\textrm{d}}\theta\, e^{-\beta W_{k}(\theta )},
  \label{eqn:functional-F}
\end{eqnarray}
Necessary conditions for the concave function $L_{k}^\star$ to
be maximal are
\begin{eqnarray}
\frac{\partial L_{k}^\star}{\partial \widetilde{A}} & = & \frac{1}{\widetilde{A}} - \int _{-\pi }^{+\pi }
{\textrm{d}}\theta\, e^{-\beta W_{k}(\theta )}=0\, ,\label{eqn:Functional-A} \\
\frac{\partial L_{k}^\star}{\partial \beta } & = &
 \frac{1}{2\beta}+  \frac{1}{2\beta_k} - U - E_{k} + \widetilde{A}\int _{-\pi }^{+\pi }{\textrm{d}}\theta\, W_{k}(\theta )\,
 e^{-\beta W_{k}(\theta )}= 0\, .\label{eqn:Functional-beta}
\end{eqnarray}
 Substituting Eq.~(\ref{eqn:Functional-A}) into Eq.~(\ref{eqn:Functional-beta}),
one gets the condition \begin{eqnarray}\label{eqn:functional-G1}
 -\frac{1}{2\beta_{k+1}} - \frac{1}{2\beta_{k}} + U + E_{k}
 -\frac{\displaystyle \int _{-\pi }^{+\pi }{\textrm{d}}\theta\ W_{k}(\theta )\,
 e^{-\beta_{k+1}W_{k}(\theta )}}{\displaystyle\int _{-\pi }^{+\pi }{\textrm{d}}\theta\,
 e^{-\beta_{k+1}W_{k}(\theta )}}=0,\label{eqn:functional-G}
\end{eqnarray}
which, since $L_k^\star$ is concave, has a unique solution. Numerically, the
solution $\beta_{k+1}$ is found by using a Newton algorithm for Eq.~(\ref{eqn:functional-G}). Then, from
Eq.~(\ref{eqn:Functional-A}), we get
 $ \widetilde{A}_{k+1} $. Finally,  we can calculate the new density
distribution from Eq.~(\ref{equationpourrhok}).

\section{Discussion of the results}
\label{sec:NumericalSol}

Using the iterative method described in the previous section, we are
able to derive the stable mass density $\rho(\theta)$ solution of
Eq.~(\ref{eqn:LE-2}) and, from that, all thermodynamic functions in the
microcanonical ensemble.  In
the first part of this section, we will show the numerical solution
obtained for $\rho(\theta)$, and its dependence on energy for a small value
of the softening parameter~$\varepsilon$.  In the second part, we will
discuss the  phase diagram of the SGR model,
both in the microcanonical and in the canonical ensemble,  when $\varepsilon$ is varied.

\subsection{Mass density, entropy and caloric curves}
\label{sec:Num-result}

For energies above a certain critical value $U_c(\varepsilon)$, the stable mass
density solution is uniform. In this case, one can compute
the entropy from Eq.~(\ref{eqn:entropy})
\begin{equation} \label{eqn:S-homo}
S=\frac{1}{2} \left (3 \log (2\pi) + 1 -\log \beta \right ) ,
\end{equation}
and the inverse temperature from Eq.~(\ref{eqn:beta-E-V})
\begin{equation}
\beta = \left( 2U - 2\overline{E}_p \right )^{-1} ,\label{formulabeta}
\end{equation}
where \begin{eqnarray}
\overline{E}_p &=&\frac{1}{2}\frac{1}{(2\pi)^2}\int_{-\pi}^{+\pi} \int_{-\pi}^{+\pi} {\rm d} \theta {\rm d} \phi\,V_\varepsilon(\theta-\phi)\\
&=&-\frac{1}{\pi\sqrt{2}} \frac{1}{\sqrt{2+ \varepsilon}}\  {\cal K} \left( {\frac{2}{\varepsilon+2}} \right ),
\end{eqnarray}
where ${\cal K}$ is the complete elliptic integral of the first kind
${\cal K} (x) \equiv \int_0^{\pi/2} {d\theta}/
 {\sqrt{1-x \sin^2 \theta}}$.

Remark that Eq.~(\ref{formulabeta}) implies that the homogenous state
cannot be continued below $U_{hom}= \overline{E}_p$, because this latter energy
corresponds to zero temperature.

\begin{figure}[ht]
\resizebox{80mm}{!}{
 \includegraphics{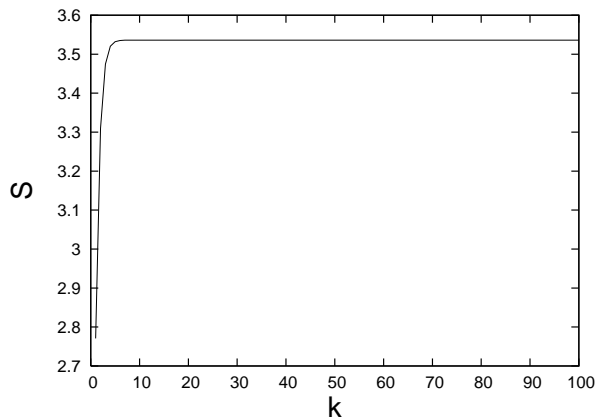}}
 \caption{Convergence of the entropy using the algorithm of
   Sec.~\ref{implementaionalgorithm} for $\varepsilon=10^{-5}, U=-1$.}
\label{fig:entropy-M2}
\end{figure}

For $U<U_c(\varepsilon)$, the stable mass distribution must be determined
numerically. We have checked in this case, that a direct iterative
method of solution of the consistency Eqs~(\ref{eqn:LE-1})
and~(\ref{eqn:LE-2}) does not always converge. On the contrary, the
novel algorithm presented in Sec.~\ref{implementaionalgorithm} ensures convergence
as shown in Fig.~\ref{fig:entropy-M2} for the entropy.

\begin{figure}[htb]
\resizebox{80mm}{!}{\includegraphics{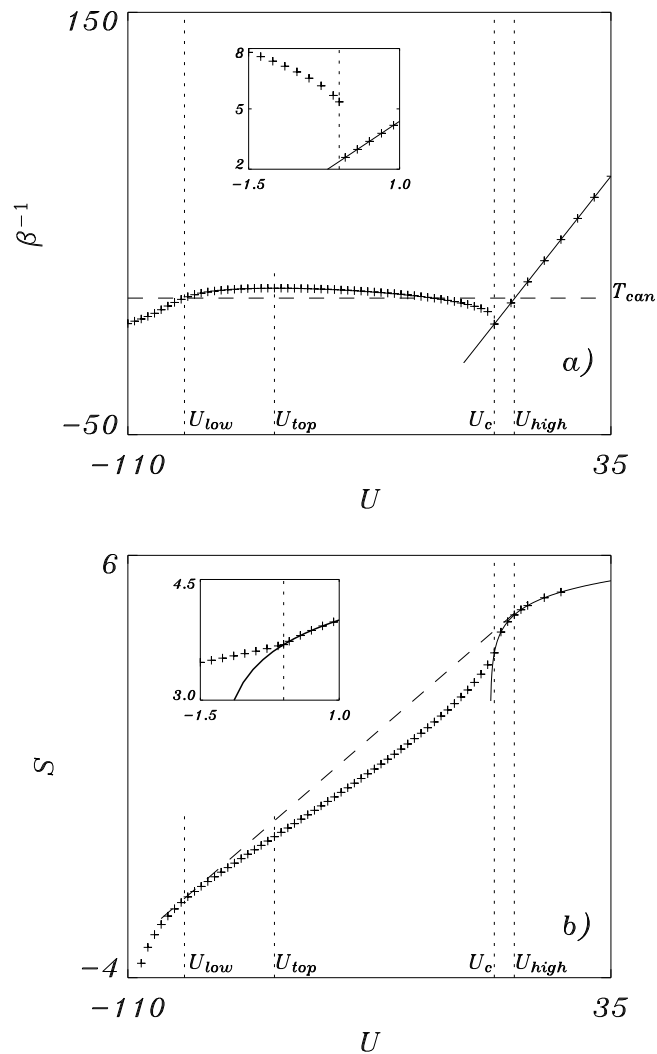}}\\
\caption{Temperature (panel (a)) and entropy (panel (b)) versus
  energy $U$ for the softening parameter value
$\varepsilon=10^{-5}$. Four values of the energy, indicated by the short-dashed vertical lines,
can be
  identified from this picture: $U_{low}\simeq-93$ and~$U_{high}\simeq6$ bound from
  below and above the  region of inequivalence of ensembles. $U_c\simeq0$ is the transition
  energy in the microcanonical ensemble. $U_{top}\simeq-66$ limits from below
  the negative specific heat region, where temperature decreases as
  energy increases. $T_{can}\simeq15$, represented with a dashed line in panel
  (a), is the canonical transition temperature and corresponds to the
  inverse slope of the entropy, both at $U_{low}$ and $U_{high}$, as
  represented by the straight dashed line in panel (b). The full
  lines represent the analytical solutions of the temperature and of
  entropy in the uniform case (see formulas ~(\ref{eqn:S-homo})
  and~(\ref{formulabeta})). They are extended slightly below $U_c$,
in the metastable phase, in order to identify them.
 The insets in panels (a) and (b) show a zoom of the
temperature and of the  entropy around $U_c$, revealing
a temperature jump at $U_c$ and different slopes of the entropy above and below $U_c$,
which emphasizes the first order nature of the phase transition.  }
\label{fig:SGR-S-e-5}
\end{figure}

\begin{figure}[ht]
 \resizebox{80mm}{!}{\includegraphics{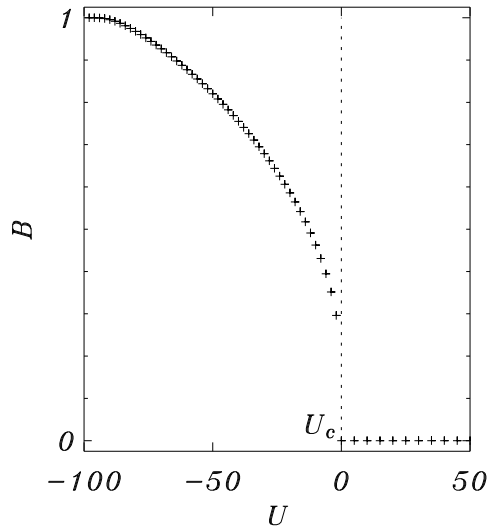}}
 \caption{``Magnetization'' $B$ versus energy $U$ for
   $\varepsilon=10^{-5}$, which emphasizes the microcanonical first order phase
   transition at $U_c\simeq0$ by showing a jump in the order parameter. }
\label{fig:SGR-Mag-e-5}
\end{figure}

In Fig.~\ref{fig:SGR-S-e-5} we show both entropy and temperature
$T=\beta^{-1}$ as a function of energy $U$. The most striking feature is
the presence of a negative specific heat region for $U_{top}\leq U\leq U_c$.
For $U_{low}\leq U\leq U_{high}$, the entropy does not coincide with its
convex envelope. Hence, microcanonical and canonical ensemble do not
give the same predictions. Indeed, the main peculiarity of the
microcanonical ensemble is that macroscopic states within this
interval are stable, while they would be either metastable or unstable
in the canonical ensemble.  The mass density is uniform above $U_c$,
while, below this value, it is localized. The appropriate order
parameter to characterize this localization is the
``magnetization''
\begin{equation}
B = \int_{-\pi}^{+\pi} {\rm d} \theta \,e^{i \theta}\, \rho(\theta) ,
\end{equation}
which vanishes if the mass distribution is uniform while it reaches
the value $B=1$ when the mass is concentrated in only one point. Intermediate
degrees of localization give intermediate values of $B$. The
``magnetization'' is plotted in Fig.~\ref{fig:SGR-Mag-e-5} as a function
of $U$. It is a decreasing function of $U$, up to $U_c$, where it has
a jump to the limiting value 0. Hence, we have here a first order
microcanonical phase transition. The first order nature of the phase
transition is confirmed zooming the entropy around $U_c$ (see the
inset in panel (b) of Fig.~\ref{fig:SGR-S-e-5}). This reveals that
this first order phase transition is of the convex-concave type (see
Ref.~\cite{Classification}).  The canonical ensemble is obtained by
taking the convex envelope of the microcanonical entropy. The
 transition is first order in the canonical ensemble and the transition
temperature $T_{can}$ is given by the inverse slope
of the entropy at $U_{low}$ and $U_{high}$. No canonical macrostate
is present in the energy range $[U_{low},U_{high}]$.

\begin{figure}[ht]
\resizebox{80mm}{!}{\includegraphics{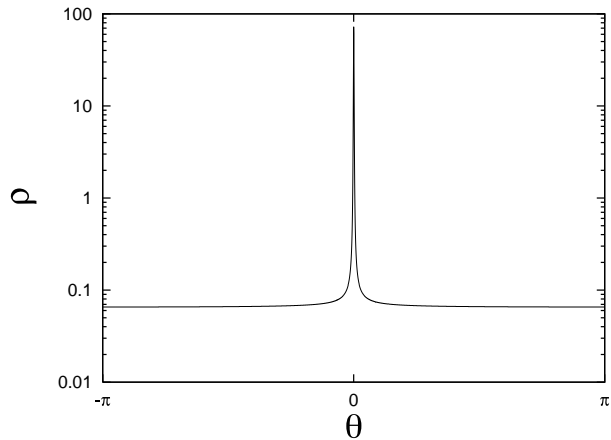}}
\caption{A typical mass density distribution $\rho(\theta)$ for
  $\varepsilon=10^{-5}$ and $U=-20.0$, in the negative specific heat region.}
\label{fig:SGR-rho20}
\end{figure}

A typical localized mass density distribution is shown in
Fig.~\ref{fig:SGR-rho20}.  It corresponds to an energy where the
specific heat is negative.

\begin{figure}[ht]
\resizebox{80mm}{!}{ \includegraphics{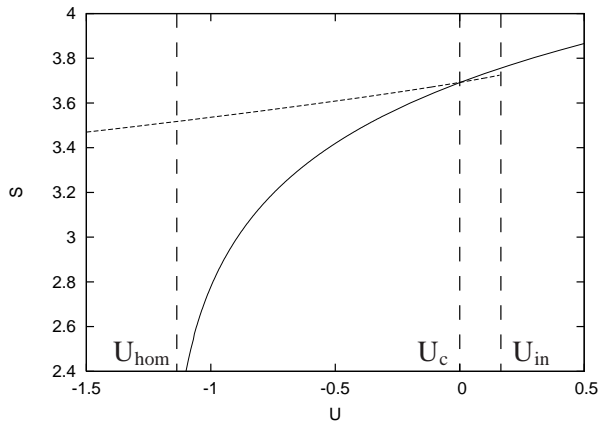}}
\caption{Both the high energy branch of the entropy versus energy
curve, corresponding to the homogeneous solution (solid line) and
the low energy branch of the inhomogeneous solution (dashed line)
are represented in this plot
 for $\varepsilon=10^{-5}$.
The two branches cross at $U_c\simeq 0$. The continuation of the
homogeneous branch into the low energy region is bounded from
below by $U_{hom}\simeq-1.19$, indicated by a vertical dashed
line. The inhomogeneous branch continues into the high energy
phase and ends at $U_{in}\simeq0.16$, again indicated by a
vertical dashed line.} \label{fig:SGR-S-meta}
\end{figure}

The first order phase transition is associated with the existence of
metastable states. Using a continuation method, we have been able to
compute them. Their entropy is represented in
Fig.~\ref{fig:SGR-S-meta} around the transition energy $U_{c}$ for the
specific case $\varepsilon=10^{-5}$.  The inhomogeneous metastable state turns
out to exist for $U_{c}\leq U\leq U_{in}$ with $U_{in}\simeq0.16$, while the
homogeneous metastable state exists for $U_{hom}\leq U\leq U_{c}$, with
$U_{hom}=\overline{E}_p\left(\varepsilon=10^{-5}\right)\simeq-1.19$.

\subsection{Behaviour as the softening parameter $\varepsilon$ is varied}
\label{sec:epsilon-dep}

\begin{figure}
\resizebox{80mm}{!}{ \includegraphics{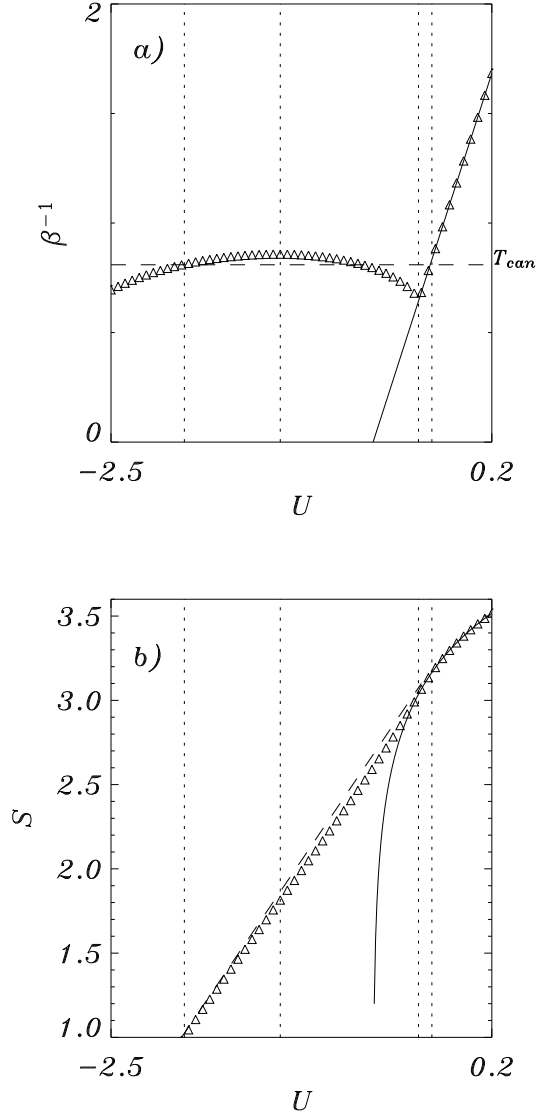}}
\caption{Panel (a): The caloric curve (triangles) for $\varepsilon=10^{-2}$.
  The dashed vertical lines indicate, from left to right, $U_{low}\simeq-1.98$, $U_{top}\simeq-1.3$,
$U_c\simeq-0.32$ and $U_{high}\simeq-0.225$.
The  homogeneous phase curve, known analytically, is shown by the continuous
line. The main difference
  with respect to Fig.~\ref{fig:SGR-S-e-5} is that now
    there is not a temperature jump at $U_c$. The phase transition is
    second order in the microcanonical ensemble, while it is still first
    order  in the canonical ensemble, at $T_{can}\simeq 0.8$. Panel (b): Entropy
    versus energy (triangles) for $\varepsilon=10^{-2}$. The entropy curve corresponding to
    the inhomogeneous distribution smoothly connects with the one of
    the homogeneous distribution (solid line). The oblique straight dashed line is tangent to the entropy at   $U_{low}$ and $U_{high}$, which delimit the energy region of ensemble inequivalence.}
\label{fig:SGR-S-e-2-det}
\end{figure}

Let us first examine a situation where the softening parameter is much
larger than previously, $\varepsilon=10^{-2}$. In the microcanonical ensemble,
Fig.~\ref{fig:SGR-S-e-2-det} shows that a concavity change still
occurs at $U_{top}\simeq-0.8$, and that a phase transition exists at
$U=U_{c}\simeq-0.3$. However, the temperature being now a continuous
function of the energy but with discontinuous derivative at $U_c$, the
phase transition is of second order, and is associated with the
symmetry breaking of the order parameter. The caloric curve shows that
this second order phase transition is of the convex-concave type.  As
it is necessary for this type of microcanonical second order phase
transition~\cite{Classification}, we observe a positive specific heat
jump at the transition point.

What we find suggests that between $\varepsilon=10^{-5}$ and $\varepsilon=10^{-2}$, there
is an intermediate value of $\varepsilon$ where a \emph{microcanonical
  tricritical point} is present. This point is signalled by two properties:
\begin{itemize}
\item The caloric curve assumes a negative infinite slope as $U$ tends to
  $U_c$ from below.
\item The upper energy of the metastable inhomogeneous phase $U_{in}$
  collapses onto $U_c$ from above, while still a continuation of the
  homogeneous phase below $U_c$ exists as an unstable phase.
\end{itemize}

\begin{figure}
\resizebox{80mm}{!}{\includegraphics{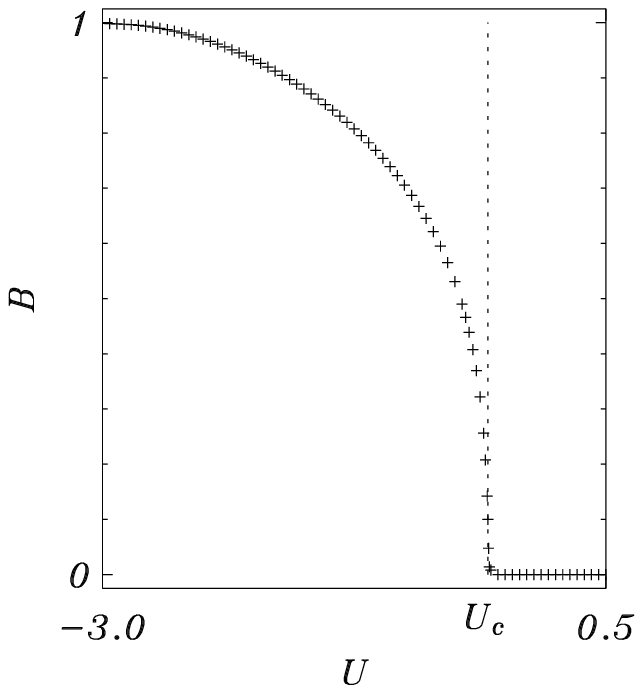}}
\caption{``Magnetization'' $B$ versus energy $U$ for
$\varepsilon=10^{-2}$ which emphasizes the microcanonical second
order phase transition at $U_c$, because the order parameter
vanishes continuously.} \label{fig:SGR-Mag-e-5bis}
\end{figure}

\begin{figure}
 \resizebox{80mm}{!}{ \includegraphics{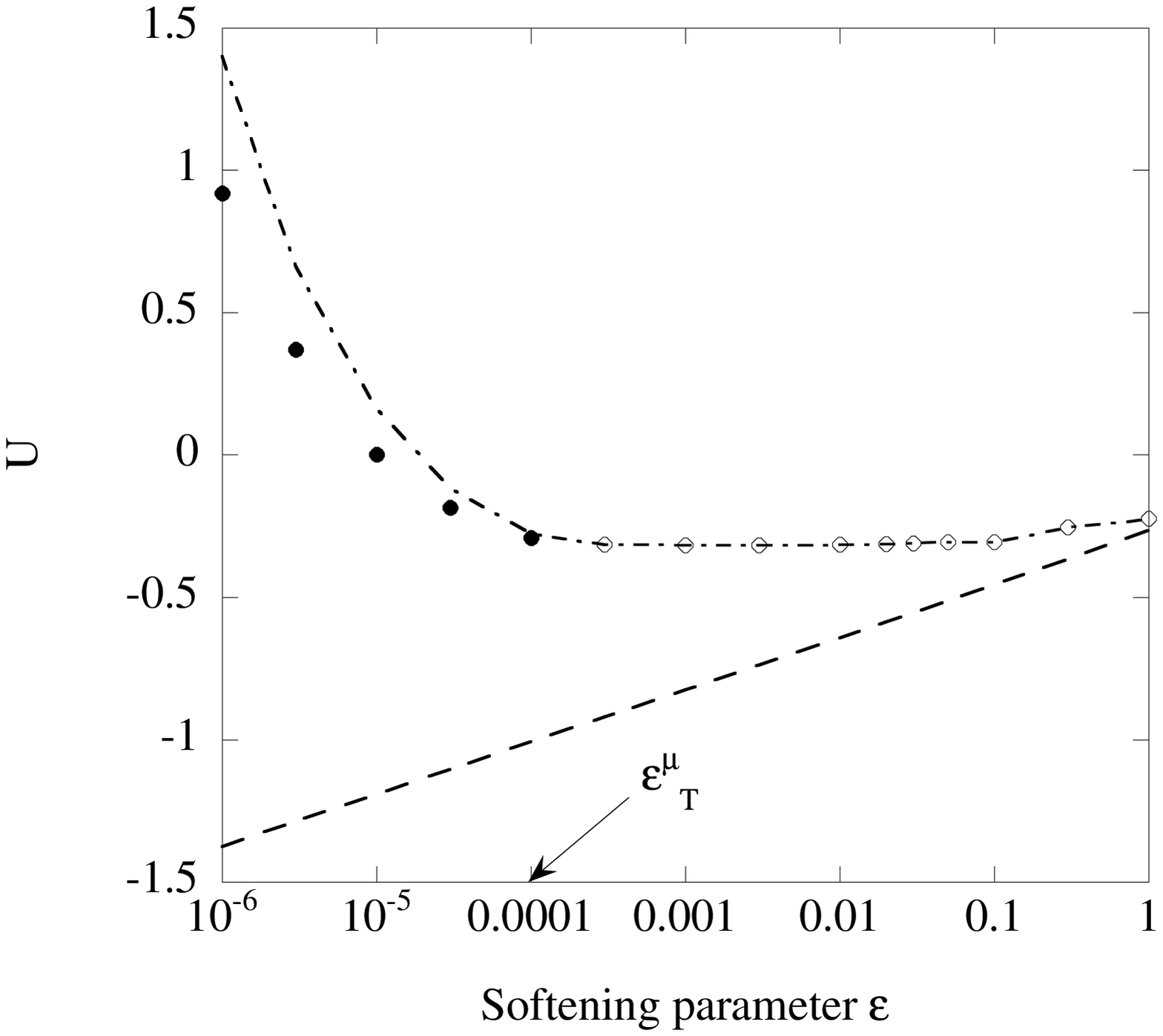}}
 \caption{The dash-dotted line represents $U_{in}(\varepsilon)$, the dashed line
   $U_{hom}=\overline{E}_p(\varepsilon)$, the filled circles the first order
   microcanonical phase transition energy and the open circles the
   second order one.  At the microcanonical tricritical point $\varepsilon_T^\mu\simeq10^{-4}$, the phase
   transition changes from first order to second order and, at the same time,
the  inhomogeneous metastable solution
   disappears.}
\label{fig:SGR-MCE}
\end{figure}

In Fig.~\ref{fig:SGR-MCE}, we have represented the $\varepsilon$-dependence of
the critical energy $U_{c}$ and of the energy bounds $U_{in}$ and
$U_{hom}$.  At the microcanonical tricritical point, $\varepsilon_{T}^\mu\simeq10^{-4}$, the end point
for the existence of the inhomogeneous metastable phase joins the
critical line. This is a generic feature of tricritical points
with symmetry breaking (see Fig. 6 of Ref.~\cite{Classification}).

\begin{figure}
\resizebox{80mm}{!}{\includegraphics{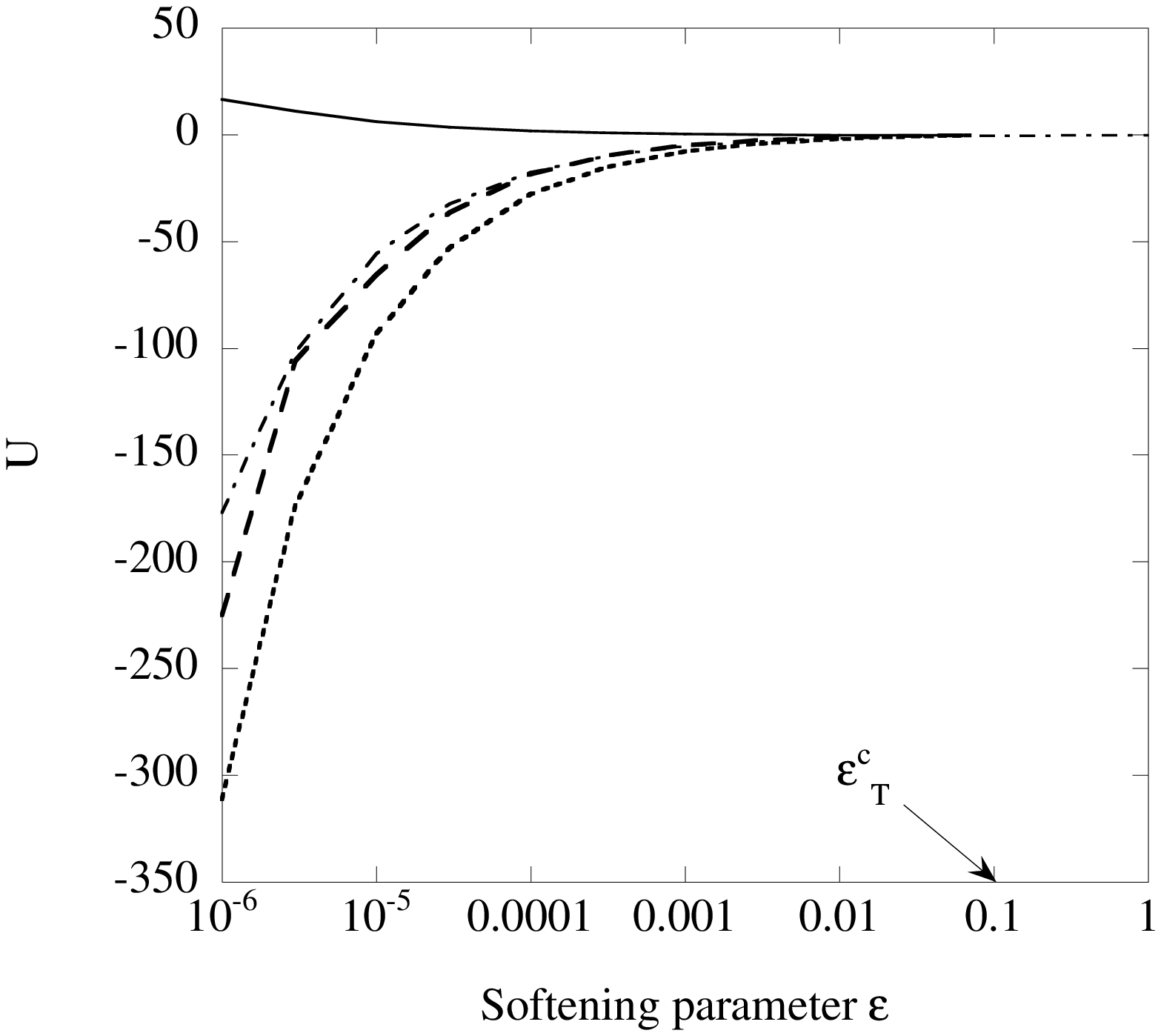}}
\caption{$\varepsilon$-dependence of $U_{low}$ (dashed line), $U_{high}$ (solid
  line) and $U_{top}$ (dash-dotted line).The canonical tricritical
  point is located at $\varepsilon_T^c\simeq 10^{-1}$ where the three curves merge.
  At this softening parameter value also the negative specific heat disappears in the
  microcanonical ensemble, while the transition becomes second order
  in the canonical ensemble. In the figure, we also show, with a
  long-dashed line, the theoretical estimate of $U_{top}^{th}\simeq -1/(4
  \sqrt{2 \varepsilon})$ obtained in Ref.~\cite{SGR-model}. }
\label{fig:SGR-CE}
\end{figure}

To locate the tricritical point in the canonical ensemble, one has
to look for the $\varepsilon$-value at which the two curves
$U_{low}$ and $U_{high}$ merge (see Fig.~\ref{fig:SGR-CE}). An
approximate estimate of this value is $\varepsilon_T^c\simeq
10^{-1}$. At the canonical tricritical point, also $U_{top}$
merges with the above curves, indicating the disappearance of the
negative specific heat region.  We thus note that ensemble
inequivalence disappears at the canonical tricritical point by the
disappearance of the inflexion point at $U_{top}$ in the entropy
curve. As it may be checked in Table I of
Ref.~\cite{Classification}, this is the only way in which ensemble
inequivalence can disappear when associated with a tricritical
point.


Summarizing, the important changes of the phase diagram of the SGR
model when $\varepsilon$ is varied are due to the existence of microcanonical
and canonical tricritical points. For $\varepsilon\leq\varepsilon_{T}^c$, there is an energy
range with ensemble inequivalence. These features were already 
observed in Refs.~\cite{Barre_Mukamel_Ruffo,Antoni_Ruffo_Torcini}.

\section{Relaxation to equilibrium}\label{sec:Evolution}

\begin{figure}
 \resizebox{60mm}{!}{\includegraphics{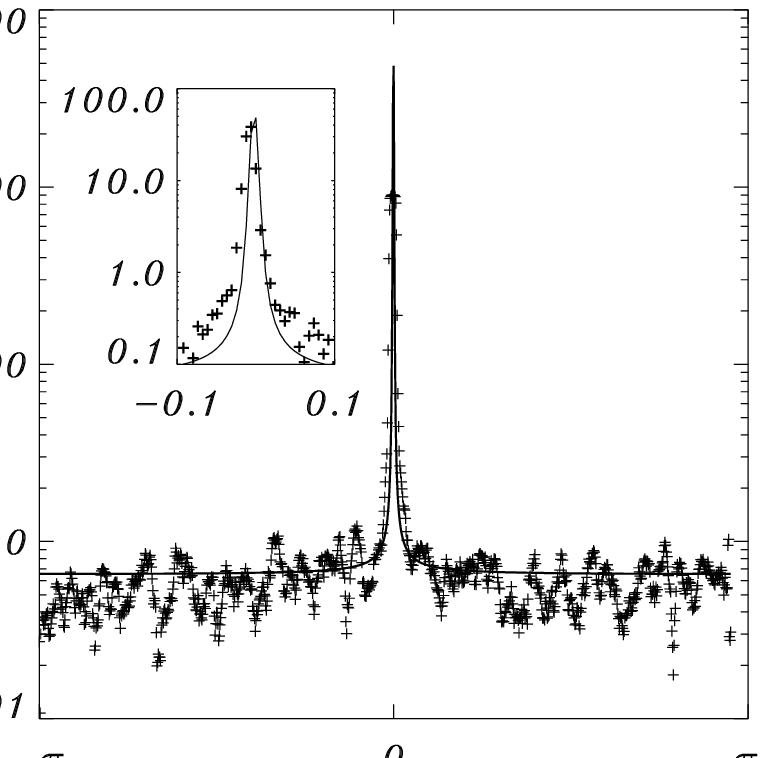}}
\vskip 0.5truecm
 \caption{Comparison of the mass density profile obtained by the iterative
method (solid line) with the result of numerical simulations (plus signs) with $N=4000$
and $\varepsilon=10^{-5}$. Parameter values
are the same of Fig~\ref{fig:SGR-rho20}. The inset is a zoom of the center of the profile.}
\label{fig:rho-CC50}
\end{figure}

We have first checked numerically if the equilibrium density profile is ever attained
in direct $N$-body simulations of Hamiltonian~(\ref{eqn:SGR-H}). In Fig.~\ref{fig:rho-CC50},
we compare the result of a numerical simulation  with
the equilibrium density profile obtained by the iterative method. The agreement is  good
in the center of the mass distribution, while the tails are still 
affected by strong finite $N$ fluctuations.
In this case,  the total energy $U=-20$
is in the region of negative specific heat and
 is well conserved using a sixth order symplectic integrator~\cite{Symplectic}.
Simulations were performed  using
GRAPE-5, a special purpose computer for gravitational force~\cite{grape5taka}.

\begin{figure}
 \resizebox{80mm}{!}{\includegraphics{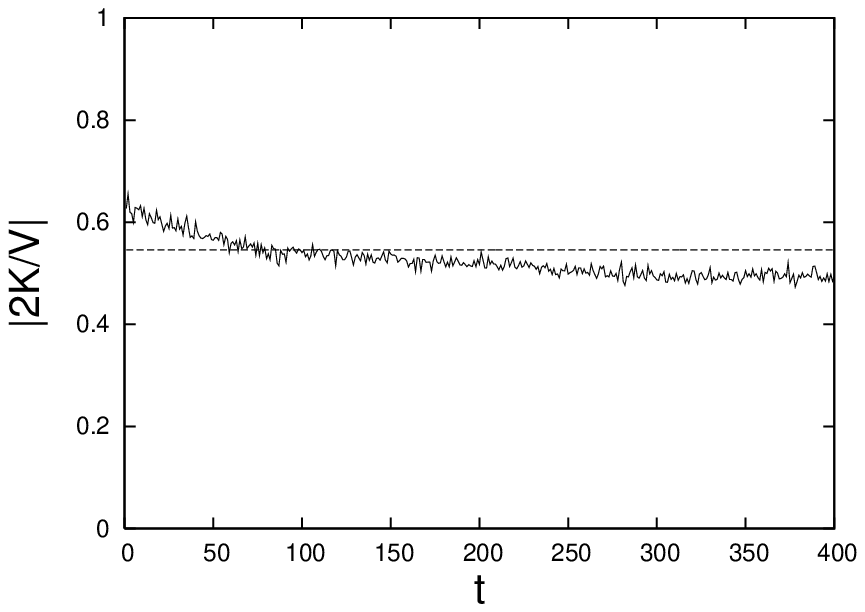}}
 \caption{Time evolution of the virial ratio $|2K/V|$ of the SGR model
 for $\varepsilon= 10^{-5}$ and $N=4000$. 
  Initially, the particles are homogeneously
distributed in the interval $[0,2 \pi/75]$ with zero kinetic energy.
The virial ratio oscillates asymptotically around the value
 0.49, which differs significantly from the equilibrium value 0.55 indicated by the dashed
horizontal line. The initial virial ratio is zero, although this time region is not visible
in the figure.}
\label{fig:Vil-CC50}
\end{figure}

However, it is well known that systems with long-range interactions display
a very slow relaxation to equilibrium~\cite{yama}. Hence, we expect that similar features
will be also exhibited by the SGR model. For instance, we can consider  a ``cold start'', where the particles are
initially homogeneously distributed on an arch ($\theta \in [\theta_{min}, \theta_{max}]$)
with zero kinetic energy.
Usually, in gravitational simulations, one looks at the evolution of the virial ratio
$|2K/V|$, which is here initially zero. The plot of the time-dependence of
this parameter is shown in Fig.~\ref{fig:Vil-CC50} for the same parameter values used previously.
One clearly observes that the system relaxes to a ``quasi-equilibrium'' state, where
the virial ratio fluctuates around a value which differs from the equilibrium one,
computed analytically. While previously,
for the mass positions  (see Fig.~\ref{fig:rho-CC50}), the relaxation was observed
on a short time scale,  we show here that a quantity
related also to velocities does not display a relaxation on the same timescale.
From previous experiences with similar cases~\cite{yama}, one expects that the relaxation should
occur on a time scale of the order of a power of $N$.

\begin{figure}
 \resizebox{160mm}{!}{\includegraphics{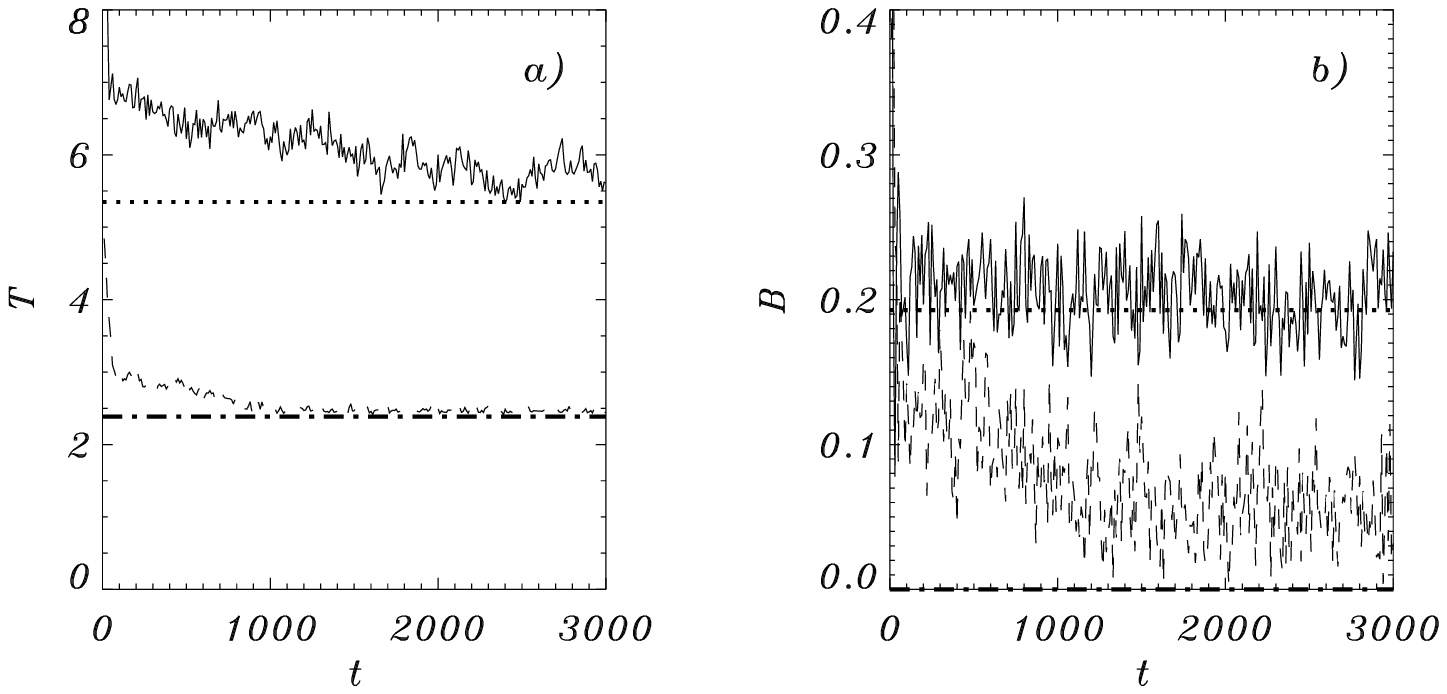}}
  \caption{Relaxation to different maximum entropy states in the SGR model for $U\simeq0$,
 $\varepsilon=10^{-5}$ and $N=10^3$.
Panel (a) shows the relaxation of the temperature either to the
inhomogeneous state value (horizontal dotted line),
 or to the homogeneous one (horizontal dash-dotted line), depending whether
the particles are initially distributed on a smaller arch $\theta \in [0, \pi/50]$ (solid line)
or a larger arch $\theta \in [0, \pi/5]$ (dashed line). In both cases,
the velocity distribution is initially a ``water bag''.
Panel (b) shows the same for the ``magnetization''.}
\label{fig:SGR-mag-U0}
\end{figure}

Even slower is the relaxation when local maxima of the  entropy
exist. This happens around the critical energy $U_c$ in the case
of a first order phase transition, e.g. for $\varepsilon=10^{-5}$.
Figs~\ref{fig:SGR-mag-U0} show the relaxation to different values
of two relevant quantities, the temperature and the
``magnetization'', when the system is initialized either with the
particles concentrated on a small arch, or on a larger one. When the
system is ``close'' to the local entropy maximum corresponding to
the clustered state, it converges to it pretty fast. The contrary
happens when the particles are more homogeneously distributed, and
then the system converges to the homogenous state. Indeed, between
the two states there is an entropy barrier which has been found to
grow as $\exp{(N)}$ for systems with long-range
interactions~\cite{torcini,chavanis}.

\section{Conclusions and perspectives}
\label{sec:conclusion}

We have fully characterized from the thermodynamic point of view
a one-dimensional model of self-gravitating particles moving on a ring~\cite{SGR-model},
which is the simplest prototype of the full
3D self-gravitating system.
Solving the equilibrium density equation by a new iterative method,
whose convergence is assured by entropy increase,
allows to derive the full phase diagram of the model both in the microcanonical
and the canonical ensemble. When the softening parameter is sufficiently small,
a negative specific heat region appears in the microcanonical ensemble, in coincidence
with the phase transition becoming first order in the canonical ensemble.
Further lowering the softening parameter, the transition becomes first order
in the microcanonical ensemble and a temperature jump appears at the transition energy.
The microcanonical and canonical tricritical points 
do not coincide~\cite{Barre_Mukamel_Ruffo}.

Dynamically, we have performed numerical experiments which
show that relaxation to equilibrium can be extremely slow. They reveal also the presence
of quasi-equilibrium states, which are ubiquitous 
in systems with long-range interactions~\cite{Dauxois02LNP}.
These states could be further characterized considering a Vlasov equation approach
as done for the HMF model in Ref.~\cite{yama}. Moreover, in the first order
microcanonical transition region a strong metastability appears and, at a given energy,
the system can relax towards different thermodynamic states.

Preliminary studies of velocity probability distributions in this model have been performed
in Ref.~\cite{SGR-model}. A similar analysis has been more recently done
for the full 3D self-gravitating system~\cite{Iguchi05}. 
In both models,non gaussians tails show up in several energy regions.
Among future directions of study of the SGR model, we think that deriving a theoretical
framework to undertand these large tails would be of particular interest. To this aim,
especially useful could be the methods developed to understand single particle diffusion 
in the HMF model~\cite{bouchet,bouchetdauxois}.

\acknowledgments
We thank Osamu Iguchi, Kei-ichi Maeda, Masahiro Morikawa,
Akika Nakamichi, and Yasuhide Sota for useful discussions.
All numerical simulation were carried out on the GRAPE system
at the Astronomical Data Analysis Center  (ADAC) of the
National Astronomical Observatory, Japan. T.T. thanks ENS Lyon for hospitality.
S.R. thanks ENS Lyon for hospitality and CNRS for financial support.
This work is part of the contract COFIN03 of the Italian
MIUR {\it Order and chaos in nonlinear extended systems},
and was partially supported by a Grant-in-Aid for Scientific
Research Fund of the Ministry of Education, Culture, Sports, Science
and Technology, Japan (Young Scientists (B) 16740152).

\end{document}